\documentclass[9pt,twocolumn,twoside,lineno]{pnas-new}
\usepackage{amsmath}
\templatetype{pnasresearcharticle} 
\usepackage[utf8]{inputenc}
\usepackage[english]{babel}
\usepackage{amsmath}
\usepackage{amsfonts}
\usepackage{amssymb}
\usepackage{graphicx}
\usepackage{float}
\usepackage{color}
\usepackage{braket}
\usepackage{sidecap}
\usepackage{mciteplus}
\usepackage{xcolor}
\usepackage{hyperref}
\newcommand{\be}{\begin{equation}}
\newcommand{\ee}{\end{equation}}
\newcommand{\ber}{\begin{eqnarray}}
\newcommand{\eer}{\end{eqnarray}}
\newcommand{\nn}{\nonumber}

\newcommand{\qv}{{\bf q}}

\def\eer{\end{eqnarray}}

\def\qv{{\bf q}}

\usepackage{fancybox}
\usepackage{bbm}
\usepackage{bm}
\usepackage{soul}

\newcommand{\vect}[1]{|{#1}\rangle}
\newcommand{\tvect}[1]{\langle{#1}|}
\newcommand{\sprod}[2]{\langle{#1}|#2\rangle}

\title{Thermal transport in compensated semimetals: a mystery explained}

\author[a,1]{Mohammad Zarenia}
\author[b]{Alessandro Principi}
\author[a]{Giovanni Vignale} 

\affil[a]{Department of Physics and Astronomy, University of Missouri, Columbia, Missouri 65211, USA.}
\affil[b]{School of Physics, University of Manchester, Oxford Road, Manchester M13 9PL, UK.}

\leadauthor{M. Zarenia} 

\significancestatement{A fundamental obstacle to the realization of efficient thermoelectric devices is the tendency of electronic thermal and electric conductivities to track each other: their ratio (the so-called Lorenz ratio) is close to having a universal value in most materials. Semimetals, having two types of electric carriers of opposite polarity, offer different modalities of conduction for heat and charge and therefore are excellent candidates to achieve anomalously large or small Lorenz ratios. Recently, it has been experimentally realized that strong interaction between carriers in a clean semimetal can significantly increase or decrease the Lorenz ratio. Here, we explain why some semimetals exhibit greatly enhanced Lorenz ratios, while others exhibit reduced ones. This has important implications for the design of thermoelectric devices.}

\authordeclaration{The authors have no competing interests.}
\correspondingauthor{\textsuperscript{1}To whom correspondence should be addressed. E-mail: zareniam@missouri.edu}

\authorcontributions{All authors contributed equally to the paper.}
\begin{abstract}
\nolinenumbers
It is well known that the electronic thermal conductivity of clean compensated semimetals can be greatly enhanced over the electric conductivity by the availability of an ambipolar mechanism of conduction, whereby electrons and holes flow in the same direction experiencing negligible Coulomb scattering as well as negligible impurity scattering. This enhancement -- resulting in a breakdown of the Wiedemann-Franz law with an anomalously large Lorenz ratio -- has been recently observed in two-dimensional monolayer and bilayer graphene near the charge neutrality point. In contrast to this, three-dimensional compensated semimetals such as WP$_2$ and Sb are typically found to show a reduced Lorenz ratio. This dramatic difference in behavior is generally attributed to different regimes of Fermi statistics in the two cases: degenerate electron-hole liquid in compensated semimetals versus non-degenerate electron-hole liquid in graphene. We show that this difference is not sufficient to explain the reduction of the Lorenz ratio in compensated semimetals. We argue that the solution of the puzzle lies in the ability of compensated semimetals to sustain sizeable regions of electron-hole accumulation near the contacts, which in turn is a consequence of the large separation of electron and hole pockets in momentum space. These accumulations suppress the ambipolar conduction mechanism and effectively split the system into two independent electron and hole conductors. We present a quantitative theory of the crossover from ambipolar to unipolar conduction as a function of the size of the electron-hole accumulation regions, and show that it naturally leads to a sample-size-dependent thermal conductivity.
\end{abstract}

\begin{document}

\maketitle
\thispagestyle{firststyle}
\ifthenelse{\boolean{shortarticle}}{\ifthenelse{\boolean{singlecolumn}}{\abscontentformatted}{\abscontent}}{}

%

%
%
\nolinenumbers
\section{Introduction}
\dropcap{T}he thermal and electric conductivities of  compensated semimetals  such as single- and double-layer graphene near the charge neutrality point have been a topic of great interest in recent years - mostly because these systems can be made very clean and feature strong Coulomb interactions between non-degenerate electron and hole carriers near the point of contact of the conduction and valence bands. This clears the way for the observation of  hydrodynamic transport, as opposed to conventional single-particle diffusive transport \cite{bandurin, ghahari, crossno, Principi_prb_2016,Narozhny_prb_2015,Briskot_prb_2015,Fritz_2008,Muller_2008, foster,svintsov2013,svintsov2018,shaffique, Principi_conductivity,lucasWF,lucasKT,zarenia,zareniaBLG,zareniaTwisted,zareniaCollapse}.

\begin{figure}[t]
\centering
\includegraphics[width=9cm]{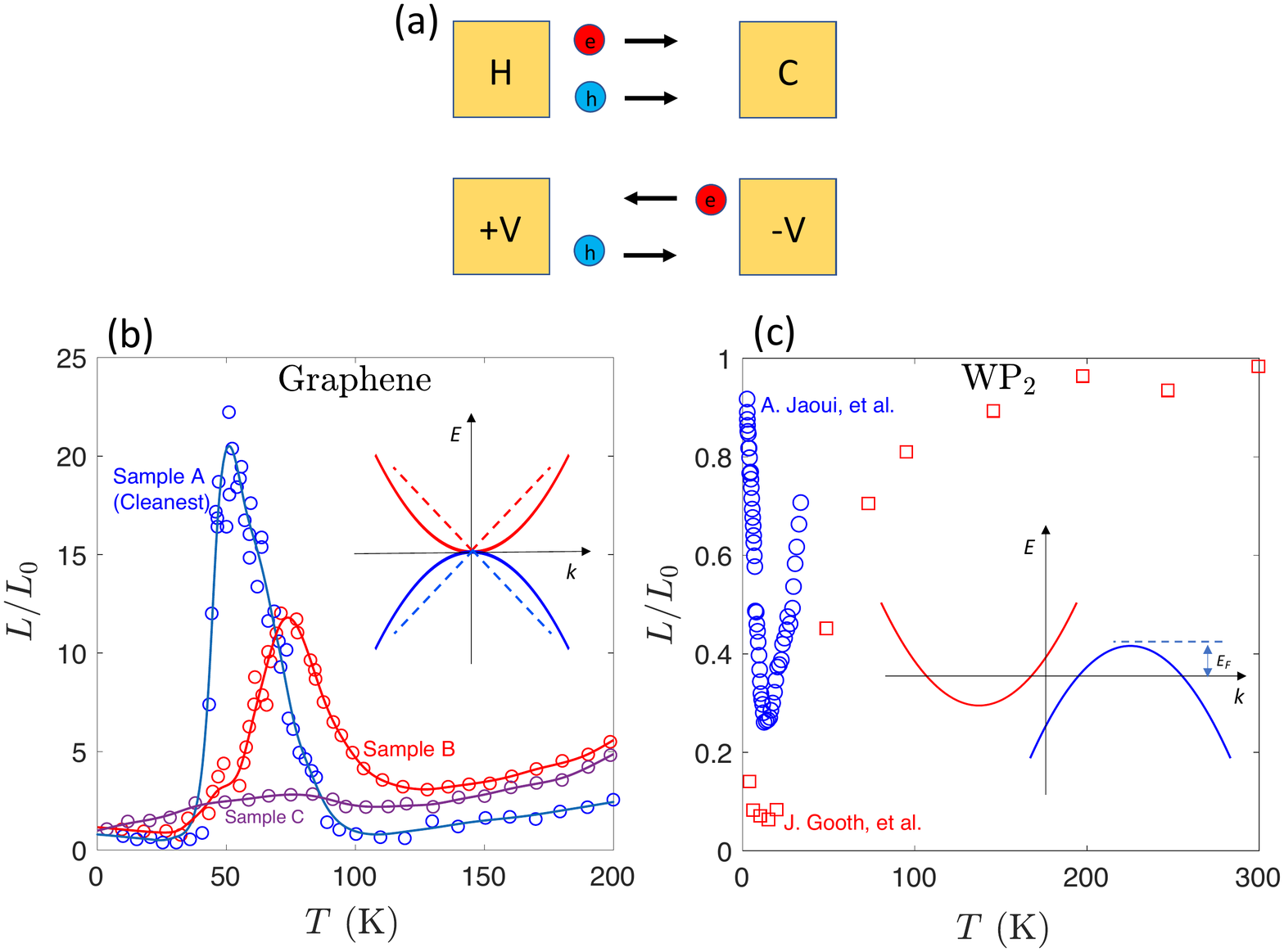}
\caption{ (a) Schematic  illustration of the difference between heat and charge current in a charge neutral system. A thermal current can be set up in a semimetal simply by letting electrons and holes drift with equal speeds in the same direction (upper row). Electric field causes electrons and holes to drift in opposite directions (lower row).
(b) Experimental observation of the enhanced Lorenz ratio in monolayer graphene (the solid lines are guides to the eye) - data were reproduced from Ref. \cite{crossno}. (c) Experimental observation of the reduced Lorenz ratio in WP$_2$ - data were reproduced from Refs. \cite{Jaoui2018, Gooth2018}.
The insets in (b) and (d) depict the low energy bands in graphene systems (dashed curves for monolayer and solid curves for bilayer) and in a compensated semimetal, respectively.  In both cases, the chemical potential is taken as the zero of the energy. } 
\label{fig:fig1}
\end{figure}

 In this regime, the thermal resistivity ($\rho_{\rm th}=\kappa^{-1}$)  -- defined under the standard condition of zero electric current -- is primarily controlled by momentum-non-conserving interactions (scattering from impurities and phonons), while  the electric resistivity ($\sigma^{-1}$) is primarily controlled by momentum-conserving electron-hole collisions. 
The physical reason for this difference is well understood.   The application of  a thermal gradient causes electrons and holes to drift in the same direction (see Fig. \ref{fig:fig1}a)   This ambipolar mode of conduction is charge-neutral and therefore automatically satisfies the condition of zero electric current, which is essential to the  measurement of the thermal conductivity.  At the same time the thermal current is directly proportional to the total momentum of the electron-hole system, which cannot be changed by electron-hole collisions. Hence, except for momentum-non-conserving processes, such as electron-impurity collisions and umklapps,  the thermal conductivity would be infinite.  On the other hand,  an electric field causes electrons and holes to drift in opposite directions (see Fig. \ref{fig:fig1}a).  Although the total momentum is now zero, electron-hole collisions transfer momentum between electrons and holes, giving rise to the Coulomb resistivity $\rho_{\rm el}$.    Under these conditions the Lorenz ratio $L=\kappa/(\sigma T)$  is  much higher than the conventional  Lorenz ratio $L_0=  (\pi^2/3) \left(k_B/e\right)^2$  (= 2.44 $\times 10^{-8}$ W $\Omega$ K$^{-2}$) of the Wiedemann-Franz  law, and is given by~\cite{zarenia}
\be\label{LorenzRatio}
L = L_0\left(1+\frac{1}{\Gamma^2}\right)
\ee
where $\Gamma^2  = (3/\pi^2) (\rho_{\rm el, dis}/\rho_{\rm el})\ll 1$ is the ratio of the ordinary Drude resistivity, $\rho_{\rm el, dis}$, to the Coulomb resistivity $\rho_{\rm el}$ -- the smaller this is, the deeper we are into the hydrodynamic regime.  (Notice that this formula is valid at or near the charge neutrality point, i.e., for chemical potential $\mu=0$ or, at least, $\mu/(k_BT)\ll \Gamma$.)
The resulting Lorenz ratio, $L>L_0$, is clearly seen in the experiments of Ref. \cite{crossno}, which we reproduce in Fig. \ref{fig:fig1}b),   and is well above what would be computed in a theory that does not take into account electron-hole scattering.

Notice that the presence of two kinds of carriers with opposite charges is essential to the enhancement of the Lorenz ratio.  If we had only one kind of carriers, then the requirement of  zero-electric current in a thermal transport experiment would  force these carriers to adopt a distribution in which their direction of  drift changes sign depending on whether their energy is above or below the Fermi level.  In this case, Coulomb interactions between the carriers would  increase the thermal resistivity, producing a Lorenz ratio that is less than $L_0$ ~\cite{vignale}, exactly the opposite of what happens in the ambipolar case.

In view of the above discussion, it may come as a surprise that well-known compensated semimetals, such as WP$_2$ do not show, experimentally, any sign of ambipolar thermal transport. On the contrary, the Lorenz ratio of this and other compensated semimetals, is found to be lower than $L_0$~\cite{behnia}, which, as we have just seen, is a signature of interaction effects in {\it unipolar} transport.  Earlier measurements on Bi~\cite{Gallo63,Uher74} also found a reduction of the Lorenz ratio rather than an enhancement.  A cartoon of the band structure of a  compensated semimetal with a negative indirect gap is shown in the inset of Fig. \ref{fig:fig1}c.  For simplicity, we assume parabolic bands of opposite curvature for electrons and holes. 
The electron and hole bands are well separated in momentum space, in contrast with those of graphene where electrons and holes coexist in the same region of momentum space.  Experimental measurements of the thermal and electric conductivity, reproduced in Fig. \ref{fig:fig1}c, clearly show the reduction of the Lorenz ratio in a range of temperatures $k_BT<\varepsilon_F$ in which electrons and holes can be safely regarded as  degenerate Fermi liquids.  Here, $\varepsilon_F$ is the Fermi energy of electrons and holes measured from the bottoms of the respective bands, while the chemical potential is $\mu=0$ as required for charge neutrality.

 
What is the reason for this difference? 

The first explanation that comes to mind invokes the different regimes of Fermi statistics of electrons and holes in the two systems.  Electrons and holes are degenerate Fermi liquids in WP$_2$, where the inverse quasiparticle lifetime scales as $(k_BT)^2/\varepsilon_F$; but, in graphene, they are non-degenerate, strongly interacting (Planckian)  particles whose inverse lifetime scales as $k_BT$.  The difference  manifests  in the behavior of the intrinsic electric resistivity (caused by electron-hole scattering): $\rho_{\rm el}$ is essentially independent of temperature in single- and double-layer graphene, but becomes proportional to $(k_BT/\varepsilon_F)^2\ll 1$ in WP$_2$.  The small value of $\rho_{\rm el}$ suggests that the ``hydrodynamicity"  parameter $1/\Gamma^2$ of Eq.~(\ref{LorenzRatio})   in  WP$_2$ is much smaller than $1$, consistent with the fact that electrons and holes are degenerate Fermi liquids.  These considerations  lead one to expect that  $L$ should be close to $L_0$, but not smaller than $L_0$.
We note that recent theoretical calculations of the thermal conductivity of compensated semimetals~\cite{Maslov2018} have yielded $L<L_0$ only because the ambipolar conduction channel was not allowed to be part of the solution.  Those results for the thermal conductivity are qualitatively similar to what would be obtained by enforcing the zero electric current conditions separately for electrons and holes, without allowing for the possibility that the the electric currents of electrons and holes cancel against each other. 

In view of the fact that the Lorenz ratio of  WP$_2$ is indeed found to be less than $L_0$ we are left with the following problem: Why is the ambipolar channel of thermal conduction apparently disabled in WP$_2$, while it is clearly operative in graphene?  In this paper we propose a resolution of this puzzle.
\begin{figure}[]
\centering
\includegraphics[width=8cm]{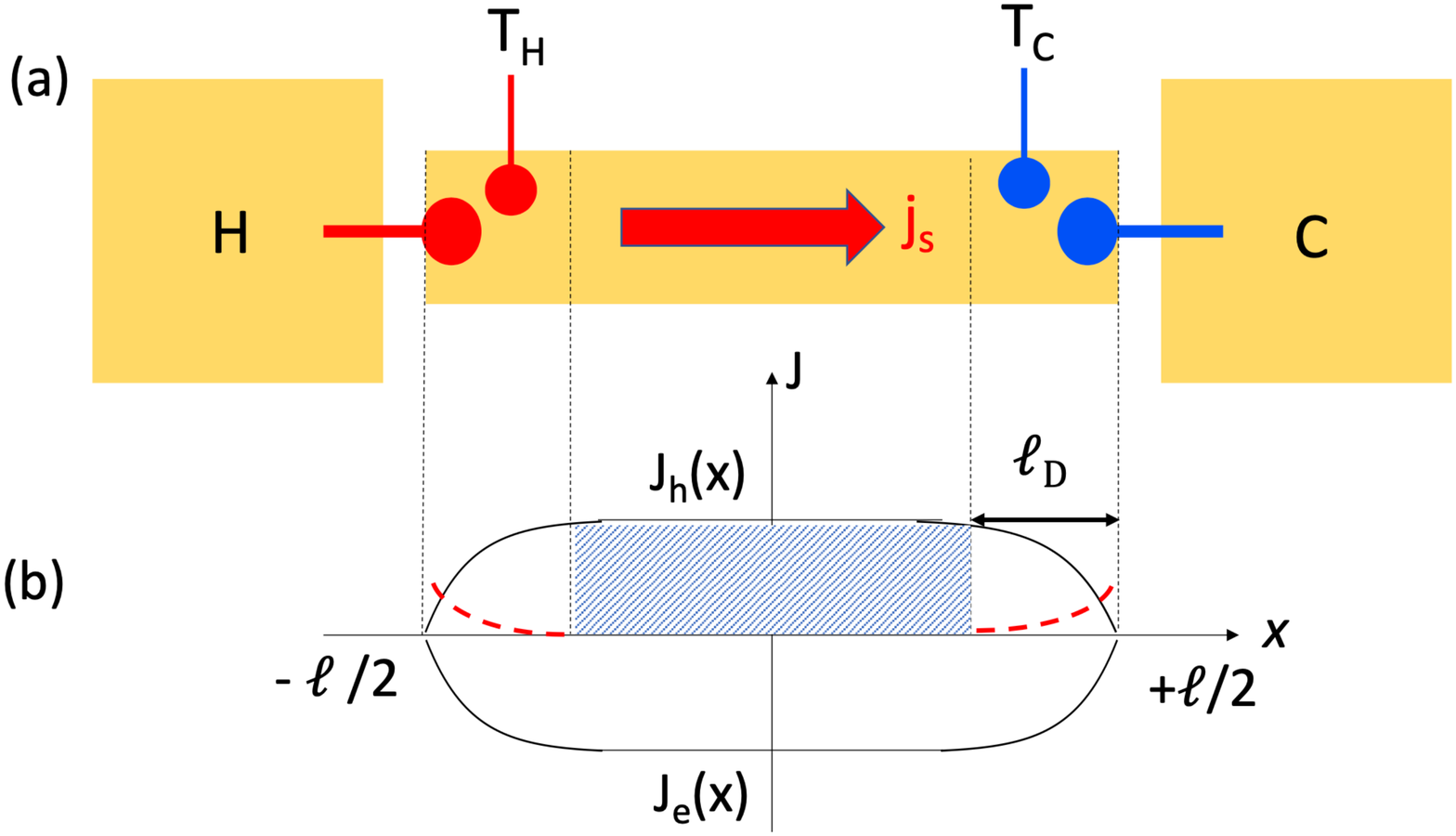}
\caption{
(a) Schematics of the thermal conductivity measurement. (b) Spatial distribution of the electron and hole components of the electric current.  $\ell_D$ is the diffusion length.  The ambipolar transport region is shaded. The red dashed lines represent the enhanced density of electrons and holes near the contacts.}
\label{fig:fig3}
\end{figure}

\section{Ambipolar transport in the presence of  contacts - qualitative description}
In a typical thermal conductivity measurement (see Fig. \ref{fig:fig3}a) no electric current is extracted from the system.  This gives us the condition
\be\label{CurrentNeutrality}
j_e+j_h=0
\ee
where $j_e$ and $j_h$ are the electric currents associated with electrons and holes respectively.   At the contacts, however, both the electron and the hole currents are expected to vanish and therefore we have the boundary condition $j_e=j_h=0$ at the contacts.  
In principle, this boundary condition could be homogeneously enforced all along the sample (we assume the sample is a channel of length $\ell$ extending from $-\ell/2$ to $+\ell/2$ along the $x$ axis).   Then the electrons and the holes  would be effectively decoupled: there would be no reason for momentum or energy to flow preferentially from one group of carriers to the other.  The thermal conductivity would be $\kappa=\kappa_e+\kappa_h$,  $\kappa_e$ and $\kappa_h$ being the thermal conductivities of electrons and holes in isolation. 
The pattern of motion would reproduce that of a system with only one kind of carrier: the drift direction would change sign depending on whether their energy is above or below the Fermi level.
Then electron-electron and hole-hole interactions would ensure that the Lorenz ratios $\kappa_e/T\sigma_e$ and $\kappa_h/T\sigma_h$, with $\sigma_e$ and $\sigma_h$ being the electric conductivities of electrons and holes in isolation, are lower than the non-interacting ratio $L_0$.  The total electric conductivity $\sigma$ is lower than $\sigma_e+\sigma_h$, due to the  effect of electron-hole collisions.  However, this effect can be neglected in the clean limit, because $\sigma_e$ and $\sigma_h$ are very large.  Therefore, under this boundary condition we would expect the Lorenz ratio to be lower than $L_0$, as it is indeed observed to be in experiments.

But why should the boundary condition $j_e=j_h=0$ be enforced homogeneously throughout the sample?  Notice that $j_e$ and $j_h$ are not separately conserved, due to the possibility of electron-hole recombination.  Therefore $j_e=0$ at the contacts does not demand $j_e=0$ everywhere.  On the contrary,  the principle of least entropy production \cite{pines} demands that the system take maximal advantage of the ambipolar channel of thermal conduction by keeping $j_e=-j_h \neq 0$ in the bulk.
The way this is achieved is by creating regions of increased electron and hole density in the vicinity of the contacts.  This is shown schematically by the red dashed lines in Fig. \ref{fig:fig3}b. The excess densities of electrons and holes are identical, so that charge neutrality is preserved,  but the local chemical potentials for electrons and holes shift in opposite direction.  The gradients of electron and hole densities act as opposing forces, which gradually bring the electron and hole currents to zero.   In the next section we will show that the size of the electron-hole accumulation regions is given by the diffusion length
\be\label{DiffusionLength}
\ell_D=\sqrt{D \tau_r}
\ee
where $D$ is the diffusion constant of electrons or holes, related to the electric conductivity by the usual Einstein relation ($D\sim v_F^2\tau$ in a degenerate Fermi liquid, $D \sim (k_BT/m)\tau$ in a non degenerate electron gas, where $\tau$ is the momentum relaxation time) and $\tau_r$ is the electron-hole recombination time. 
Notice that $\ell_D$ can be very large in a clean semimetal with a long electron-hole recombination time. For example, with a diffusion constant on the order of 10$^4$ cm$^2$/s and an electron-hole recombination time on the order of 10$^{-6}$ s (e.g. see chapter 4  in \cite{Streetman}) we obtain $\ell \simeq 10^{-1}$ cm which is comparable to the size of experimental samples  \cite{Jaoui2018, Gooth2018}. It is  also worth noting that this mechanism of gradual suppression of the current is unique to ambipolar systems.  In a unipolar system, carrier accumulation is inevitably associated with charge accumulation and the diffusion length is replaced by the much smaller screening length: the electric current is suppressed all over the sample by the uniform electric field generated by a surface charge layer. 

The following qualitative picture emerges from our discussion.  In a typical thermal conductivity measurement the system splits into three sections: (i) A central section of length $\ell-2 \ell_D$ (assuming $\ell>2\ell_D$) in which thermal transport  occurs via the ambipolar channel with $j_e=-j_h \neq 0$ and the thermal resistivity is given by $\rho_{\rm th,ambi}$  (ii) Two accumulation regions of length $\ell_D$ adjacent to the contacts, in which $j_e$ and $j_h$ are essentially zero and the thermal conductivity is given by $\rho_{\rm th,uni}=(\kappa_e+\kappa_h)^{-1}$.  The thermal resistivities of the three sections add in series, leading us to the final result
\be\label{QualitativeResult}
\rho_{\rm th} = \alpha (\ell) \rho_{\rm th,uni}+[1-\alpha(\ell)]\rho_{\rm th,ambi}
\ee
where $\alpha (\ell) \simeq 2 \ell_D/\ell$ for $\ell>2 \ell_D$, and $\alpha (\ell) =1$ for $\ell<2 \ell_D$.  This qualitative result will be substantiated in the next section by detailed calculations.  In particular, we will derive a more accurate form of the weight function
\be \label{WeightFunction}
\alpha(\ell) = \frac{2 \ell_D}{\ell}\tanh \left( \frac{\ell} {2 \ell_D}\right)\,. 
\ee
If, as we expect in very clean samples, $\rho_{\rm th,ambi} \ll \rho_{\rm th,uni}$ Eq.~(\ref{QualitativeResult}) can be further simplified to
 \be\label{QualitativeResultSimplified}
\rho_{\rm th} = \alpha (\ell) \rho_{\rm th,uni}\,.
\ee 
Here the  dependence of the thermal resistivity on the sample size along the direction of flow is evident -- as well as a distinct possibility to get $\rho_{\rm th}\simeq \rho_{\rm th,uni}$ 
when $\ell$ and $2\ell_D$ are comparable. 
No such complications arise in measurements of the electric conductivity, because the system remains homogeneous in those measurements.

According to this description, the  difference between compensated semimetals like WP$_2$ and graphene arises from the difference between their electron-hole equilibration times.  In WP$_2$  electrons and holes are well separated in momentum space, making the recombination process very slow. As a result, the diffusion length becomes comparable to the size of the sample and the thermal conductivity reduces to the sum of the thermal conductivities of electrons and holes in isolation, implying a Lorenz ratio lower than $L_0$, as discussed above. In single- and double-layer graphene, electrons and holes coexist in the same (small) region of momentum space.  Transfer of non-equilibrium carriers between the conduction and valence bands is  fast, preventing the establishment of different local chemical potential for electrons and holes. Therefore the diffusion length is negligible and the thermal resistivity plummets, leading to a Lorenz ratio higher than $L_0$. 

Throughout this paper we have assumed that the current density remains uniform in the direction {\it perpendicular} to the flow.  Thus, we have deliberately disregarded contributions to the resistances arising from the transverse electronic viscosity and boundary conditions which mandate the vanishing of the electronic current along the lateral boundaries of the channel. 
This corresponds to considering a wide conduction channel.  A detailed analysis of narrow-channel effects is beyond the scope of this paper.

\section{Ambipolar transport in the presence of  contacts - quantitative theory}
\label{sect:ambipolar_quantitative_theory}
In this section we derive the $2\times2$ matrix of thermoelectric resistivities for a 1D channel ($-\ell/2\leq x\leq \ell/2$). The latter relates electric and thermal currents to electric fields and thermal gradients. To simplify the following derivation, we now define $\vect{j_{ns}} = {}^t(j_n, j_s)$ and $\vect{F_{ns}} = {}^t(-e E, -k_{\rm B}\partial_x T)$ the vectors of thermoelectric currents and fields, respectively. Here $j_n = j_e+j_h$ and $j_s$ are the electric and thermal currents, respectively, while $E$ is the electric field and $\partial_x T$ the temperature gradient. Then, the resistivity matrix ${\hat \rho}$, such that $\vect{F_{ns}} = {\hat \rho} \vect{j_{ns}}$, has the form
\be \label{eq:rho_2by2_def}
{\hat \rho}  = \left(
\begin{array}{cc}
\rho_{\rm el} + Q^2 \rho_{\rm th} & - Q\rho_{\rm th}
\\
- Q\rho_{\rm th} & \rho_{\rm th}
\end{array}
\right)
~,
\ee
where $\rho_{\rm el}$ and $\rho_{\rm th}$ are the electric and thermal resistivities in reduced units. That is to say, they are the usual electric and thermal resistivities multiplied by $e^2$ and $k_{\rm B}^2 T$, respectively, while $Q$ is the Seebeck coefficient in units of $k_{\rm B}/e$. Throughout this paper we work with these reduced units.

This well-establish two-mode description is however insufficient in describing thermoelectric transport in systems where conduction can occur via both electrons and holes, if one wishes to separately impose boundary conditions on the particle and hole currents $j_e$ and $j_h$. It is in fact clear that, by its own construction, such description allows imposing boundary conditions only on the total electric current, $j_n$, which is the sum of  electron and hole currents. To treat these currents separately, it is  necessary to extend this theory by adding a third mode, the ``imbalance'' current $j_\delta = j_e - j_h$, as well as the corresponding imbalance field $F_\delta = -\partial_x(\mu_e - \mu_h)$. Here, $\mu_e$ and $\mu_h$ are the electron and hole chemical potentials, respectively. Indeed, by taking linear combinations of the imbalance and electric currents, it becomes possible to separately describe the propagation of electrons and holes.

We stress that the imbalance mode plays a rather special role in the present theory. From an experimental perspective, only two fields and currents, the electric and thermal ones, can be externally applied and measured. On the contrary, $j_\delta$ and $F_\delta$ are not directly accessible. They  represent the internal rearrangement that the particle flow undergoes as a result of the application of external probes, while being subject to the boundary conditions. Their presence in the theory is vital to the correct implementation of boundary conditions and particle-hole recombination.  However, in order to describe experiments, it is sufficient to down-fold such unfamiliar three-mode theory, resulting from the introduction of imbalance currents and fields, to the more conventional two-mode one of Eq.~(\ref{eq:rho_2by2_def}).
Here we show that, by applying the boundary conditions on $j_\delta$ in the presence of particle-hole recombination, we are able to integrate out the imbalance current and reduce the three-mode thermoelectric resistivity to the more familiar $2\times2$ matrix of Eq.~(\ref{eq:rho_2by2_def}). From that we will then be able to read out the values of electric and thermal resistivities, as well as of the Wiedemann-Franz ratio and the Seebeck coefficient. 

In the three-mode theory, the fields are related to the currents via a $3\times 3$ resistivity matrix:
\be \label{eq:rho_split_def}
\left(
\begin{array}{c}
\vect{F_{ns}}
\\
\hline
F_\delta
\end{array}
\right)
=
\left(
\begin{array}{c|c}
{\hat \rho}_{ns} & \vect{\rho_\delta}
\\
\hline
\tvect{\rho_\delta} & \rho_{\delta\delta}
\end{array}
\right)
\left(
\begin{array}{c}
\vect{j_{ns}}
\\
\hline
j_\delta
\end{array}
\right)
~.
\ee
Here ${\hat \rho}_{ns}$ is a $2\times 2$ block, whereas $\vect{\rho_\delta}$ is a two-component vector. Hereafter $\tvect{v}$ denotes the transposed of the vector $\vect{v}$. The vector of currents on the right-hand side of Eq.~(\ref{eq:rho_split_def}) specifies the state in which the system is prepared. Once such state is defined, this equation tells us which potential drops, thermal gradients and imbalance fields can be measured at the boundaries of the sample. We note that the specific forms of ${\hat \rho}_{ns}$, $\vect{\rho_\delta}$ and $\rho_{\delta\delta}$ are immaterial, as we proceed to show. The only property of the $3\times 3$ matrix of Eq.~(\ref{eq:rho_split_def}) that we will use in what follows is that its determinant vanishes. We stress that such property is not generic to all thermoelectric matrices, but is a consequence of the existence of a conserved mode (the total momentum) in the present theory.

In fact, when electron-electron interactions are the dominant scattering mechanism, and barring Umklapp processes, the total momentum is a conserved quantity which must always be included in the theory, regardless of boundary conditions. This can be accomplished in two ways. One possibility is that the total momentum is already present explicitly in the $3\times3$ resistivity matrix of Eq.~(\ref{eq:rho_split_def}). This happens in very specific cases in which one of the three currents ($j_n, j_s$ or $j_\delta$) coincides with the momentum density. For example, in a parabolic band electron gas the electric current density coincides with the momentum density, whereas for massless Dirac fermions (e.g., in undoped graphene) the momentum density is directly proportional to the heat current density. Finally, in a gapless parabolic-band semimetal such as undoped bilayer graphene, the momentum density coincides with the imbalance current density $j_\delta$. 

In all these cases, the current that is proportional to the momentum {\it cannot} decay over time, since particle-particle collisions do not affect it. Once launched, it can only be relaxed by momentum-non-conserving scattering processes (e.g., electron-phonon collisions). By the very definition of hydrodynamic regime of transport, however, such processes seldom occur and are in fact neglected altogether in a first approximation. This fact has a striking consequence. If the system is prepared in a state in which only such conserved current exists, since it experiences neither resistance nor dissipation during its propagation, it {\it cannot} give rise to a drop in electric field or thermal gradient. Mathematically, if such {\it nontrivial} state is introduced on the right-hand side of Eq.~(\ref{eq:rho_split_def}), and is thus multiplied by the $3\times3$ resistivity matrix, it produces a null vector of fields. It is, therefore, a ``zero mode'' of the resistivity matrix. Since it is nontrivial, {\it i.e.} it is not the vector with all currents equal to zero, this in turn implies that the determinant of the $3\times3$ resistivity matrix of Eq.~(\ref{eq:rho_split_def}) must vanish.

In general, however, none of the three currents coincides with, or is directly proportional to the total momentum. Therefore, to include such mode one should in principle start from four-mode theory, the fourth component being the momentum. Then, via a down-folding procedure similar to the one we will describe momentarily, one can obtain the $3\times3$ resistivity matrix of Eq.~(\ref{eq:rho_split_def}). This procedure is shown in App.~\ref{app:Nm1_Nm1_matrix}: the end result is that the momentum mode is {\it implicitly} included in the $3\times3$ resistivity matrix and its determinant still vanishes (it is indeed possible to construct a current which is a combination of particle, thermal and imbalance ones that cannot decay over time). As we proceed to show, the vanishing of the determinant of the three-mode matrix, consequence of the presence of a conserved mode in the theory, plays a fundamental role in describing the transition between unipolar and ambipolar regimes, as well as the size-dependence of the thermal resistivity.

To take into account particle-hole recombination and boundary conditions in the thermoelectric transport, we now assume that the imbalance density, $n_\delta$, satisfies the following continuity equation:
\be \label{eq:main_recomb_1}
\partial_t n_\delta + \partial_x j_\delta = -\frac{\nu_0}{\tau_r} (\mu_e-\mu_h)
~,
\ee
where $\tau_r$ is the electron-hole recombination time, while $\nu_0$ is the density of states of electrons and holes (assumed to be equal) at the Fermi level.  Eq.~(\ref{eq:main_recomb_1}) can be derived from the Boltzmann equation. In general, the collision integral of electron-electron interactions does not conserve the imbalance density and therefore the latter decays over time with a typical time scale $\tau_r$. Note that conservation of the imbalance density is obtained in the limit $\tau_r \to \infty$, hence Eq.~(\ref{eq:main_recomb_1}) is completely general. 


Before continuing, it is necessary to discuss which boundary conditions apply in different situations. In a typical measurement of the thermal resistivity, the channel is connected to two thermal reservoirs. There is no charge transfer to the reservoirs and only heat can be exchanged between them and the channel. Hence, the currents of electrons and holes have to vanish at the boundaries. This in particular implies that $j_\delta(\pm \ell/2) = 0$. This leads to the accumulation of electrons and holes at the boundaries. Such accumulation is required to stop the two currents from propagating in the channel. Hence, the imbalance field $F_\delta$ can be finite. On the contrary, when the electric resistivity is measured, a charge current is passed through the system and a voltage drop is detected. In this case, the imbalance current needs not to vanish at the boundaries and is in fact uniform throughout the channel. However, since there is no applied imbalance field, $F_\delta$ must vanish.

We will start by considering the measurement of the thermal resistivity.
Taking the derivative of Eq.~(\ref{eq:main_recomb_1}), in the steady state we get
\be \label{eq:main_recomb_2}
\partial_x^2 j_\delta(x) = \frac{\nu_0}{\tau_r}  F_\delta
~.
\ee
From the last line of Eq.~(\ref{eq:rho_split_def}), we get that $F_\delta = \sprod{\rho_\delta}{j_{ns}} + \rho_{\delta\delta} j_\delta$. Using this into Eq.~(\ref{eq:main_recomb_2}), and then solving by imposing the boundary conditions $j_\delta(\pm \ell/2) = 0$, we get 
\be \label{eq:j_delta_channel_sol}
j_\delta(x) = -\frac{\sprod{\rho_\delta}{j_{ns}}}{\rho_{\delta\delta}} \big[ 1 - {\alpha}(\ell, x) \big]
~.
\ee
In this equation,
\be
{\alpha}(\ell, x) = \frac{\cosh(x/\ell_D)}{\cosh\big[\ell/(2\ell_D)\big]}
~,
\ee
where $\ell_D \equiv \sqrt{\tau_r/(\nu_0\rho_{\delta\delta})}$ is the recombination length. To get Eq.~(\ref{eq:j_delta_channel_sol}), we have assumed that the thermal and particle currents are constant throughout the channel, while the (electric, thermal and imbalance) fields depend on position. This implies that there is no loss of energy along the channel. This is compatible with the system being in the hydrodynamic regime: energy loss occurs via phonon emission, which is however assumed to occur at a much slower rate than electron-electron collisions. 

We note that the hydrodynamic hypothesis also explains why $\ell_D$ can assume drastically different values in compensated semimetals and in, e.g., graphene systems. In a compensated semimetal, electron and hole Fermi surfaces are centered at distant points of the Brillouin zone. Electron-hole recombination occurs at the Fermi surface and requires a large transfer of momentum, much larger than the typical Fermi momenta of the involved particles. Therefore, it requires momentum-non-conserving scattering process to be effective in equilibrating particles and holes with each other. But this is exactly what is prevented in the hydrodynamic regime of transport, in which momentum-non-conserving collisions with impurities or phonons seldom occur. Hence, $\tau_r$ becomes very large, while electrons and holes are largely independent. When the recombination time $\tau_r\to \infty$, $\ell_D$ diverges and ${\alpha}(\ell, x) \to 1$. Under this condition, the imbalance current of Eq.~(\ref{eq:j_delta_channel_sol}) vanishes everywhere and the system behaves as two independent unipolar systems. 

On the contrary in, e.g. graphene systems, electron-hole recombination (and therefore the equalization of their chemical potentials) occurs at a much faster rate, with typical time scales of few tens of femtoseconds. Hence, the typical relaxation times for imbalances in chemical potential are very short, {\it i.e.} $\tau_r\to 0$. In this case, $\ell_D$ vanishes and ${\alpha}(\ell, x) \to 0$. Since the imbalance current can be finite, the system displays  ambipolar behavior. We stress that, in graphene, electron-hole recombination occurs in general much faster than cooling, which has typical time scales of few picoseconds \cite{Chen2019}.

Substituting the result of Eq.~(\ref{eq:j_delta_channel_sol}) into the first line of Eq.~(\ref{eq:rho_split_def}), we obtain an equation of the form $\vect{F_{ns}} = {\tilde \rho}(x) \vect{j_{ns}}$, where the $2\times 2$ position-dependent thermoelectric matrix is
\be \label{eq:tilde_rho_ns}
{\tilde \rho}(x) = {\hat \rho}_{ns}  - \frac{\vect{\rho_\delta}\tvect{\rho_\delta}}{\rho_{\delta\delta}} \big[ 1 - {\alpha}(\ell, x) \big]
~.
\ee
According to the discussion above, from Eq.~(\ref{eq:tilde_rho_ns}) we can define the unipolar and ambipolar resistivity matrices as
\ber
&&
{\hat \rho}_{\rm uni} \equiv \lim_{{\alpha}(\ell, x) \to 1} {\tilde \rho}(x) =  {\hat \rho}_{ns}
~,
\nn\\
&&
{\hat \rho}_{\rm ambi} \equiv \lim_{{\alpha}(\ell, x) \to 0} {\tilde \rho}(x) = {\hat \rho}_{ns}  - \frac{\vect{\rho_\delta}\tvect{\rho_\delta}}{\rho_{\delta\delta}} 
~.
\eer
Each infinitesimally thin slice of the channel at position $x$ contributes a resistivity ${\tilde \rho}(x)$, which is in series to those of all other slices. The total resistivity of the channel is therefore obtained by summing the resistivities of the infinitesimally thin slices that compose it, and dividing the result by its total length $\ell$. This is equivalent to averaging Eq.~(\ref{eq:tilde_rho_ns}) over the length of the channel. We finally obtain the sought $2\times 2$ thermoelectric matrix subject to thermal-measurement boundary conditions
\begin{equation} \label{eq:WF_rhons_1}
{\hat \rho} = {\hat \rho}_{\rm uni} \alpha(\ell)  + {\hat \rho}_{\rm ambi} \big[1 - \alpha(\ell)\big]
~.
\end{equation}
Note that we can easily add a contribution due to momentum-non-conserving processes ${\hat \rho}_{\rm D}$ in series to ${\hat \rho}$ by replacing ${\hat \rho}_{\rm uni} \to {\hat \rho}_{\rm uni} +{\hat \rho}_{\rm D}$ and ${\hat \rho}_{\rm ambi} \to {\hat \rho}_{\rm ambi} +{\hat \rho}_{\rm D}$. Comparing Eq.~(\ref{eq:WF_rhons_1}) with the definition~(\ref{eq:rho_2by2_def}), we immediately identify the thermal resistivity
\be \label{eq:theory_rho_t_result}
\rho_{\rm th} = \alpha(\ell) \rho_{\rm th, uni}  + \big[1 - \alpha(\ell)\big] \rho_{\rm th, ambi}
~.
\ee
Using Eqs. (\ref{eq:rho_split_def}) and (\ref{eq:WF_rhons_1}) we can also derive  the Seebeck coefficient as
\be \label{eq:theory_Q_result}
Q = \frac{\alpha(\ell) \rho_{\rm th, uni} Q_{\rm uni}  + \big[1 - \alpha(\ell)\big] \rho_{\rm th, ambi} Q_{\rm ambi}}{\alpha(\ell)\rho_{\rm th, uni}  + \big[1 - \alpha(\ell)\big] \rho_{\rm th, ambi}}
~,
\ee
where $Q_{\rm uni}$ and $Q_{\rm ambi}$ are the Seebeck coefficients of the system in the unipolar and ambipolar regimes, respectively.
\begin{figure}[]
\includegraphics[width=8cm]{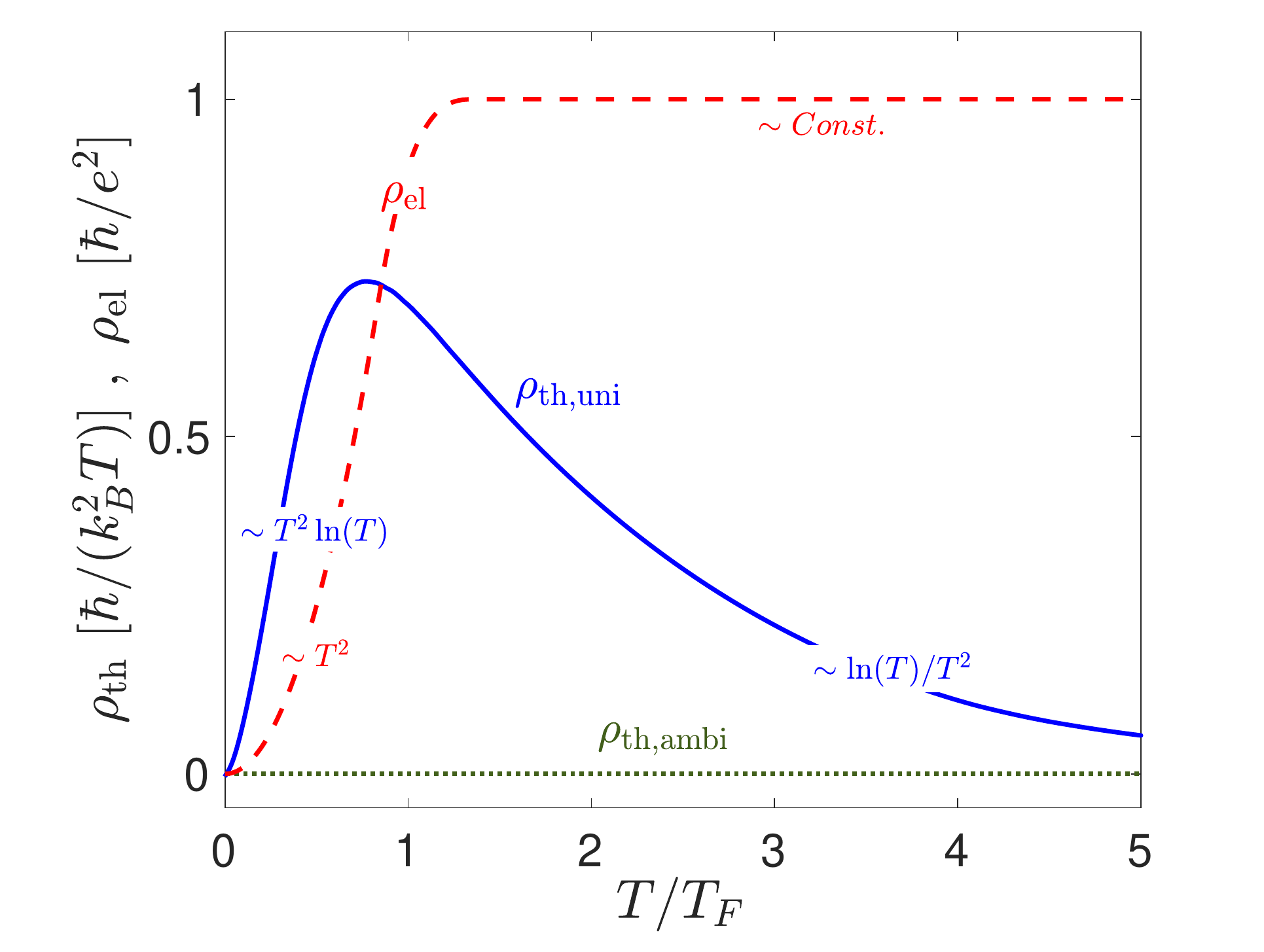}
\caption{Intrinsic unipolar and ambipolar thermal resistivities (as defined in the text) as well as the intrinsic electric resistivity  as functions of temperature (scaled with $T_F$). In the presence of disorder the total electric and thermal resistivities, respectively, become $\rho_{el}+\rho_{\rm el,dis}$ and $\rho_{\rm th, uni/ambi}+\rho_{\rm th,dis}$ (not shown in the figure), where $\rho_{\rm el,dis}$ is constant at low $T$ (impurity-dominated regime) and linearly scales with $T$ when phonons are relevant. Making use of the Wiedemann-Franz law for a  non-interacting disordered system,  we have assumed $\rho_{\rm el,dis}/\rho_{\rm th,dis}=\pi^2/3$ (in reduced units).}
\label{fig:RthRel}
\end{figure}

To derive the electric resistivity, we have to start again from Eq.~(\ref{eq:rho_split_def}) and apply the boundary condition $F_\delta = 0$. Since we impose no condition on the imbalance current, the latter has a uniform value which is determined by the applied fields. In this case, the electric resistivity is simply that of the ambipolar channel, {\it i.e.} $\rho_{\rm el} = \rho_{\rm el, ambi}$. In fact, given that the boundary conditions do no treat particles and holes separately, there is no such thing as a ``unipolar'' electric resistivity. This can be seen also mathematically, by setting $F_\delta = 0$ in Eq.~(\ref{eq:rho_split_def}) and solving its last line. The result is $j_\delta = -\sprod{\rho_\delta}{j_{ns}}/\rho_{\delta\delta}$, which is identical to Eq.~(\ref{eq:j_delta_channel_sol}) in the limit ${\alpha}(\ell, x) \to 0$. Therefore, as expected, the system behaves as purely ambipolar and ${\hat \rho} = {\hat \rho}_{\rm ambi}$. Therefore, the Lorenz ratio reads
\be \label{eq:theory_WF_result}
L=\frac{\rho_{\rm el}}{\alpha(\ell) \rho_{\rm th, uni}  + \big[1 - \alpha(\ell)\big] \rho_{\rm th, ambi}}
~.
\ee
In the absence of interactions and in the unipolar regime, $L$ tends to $L_0$, the value prescribed by the Wiedemann-Franz law.
\begin{SCfigure*}[]
\includegraphics[width=12cm]{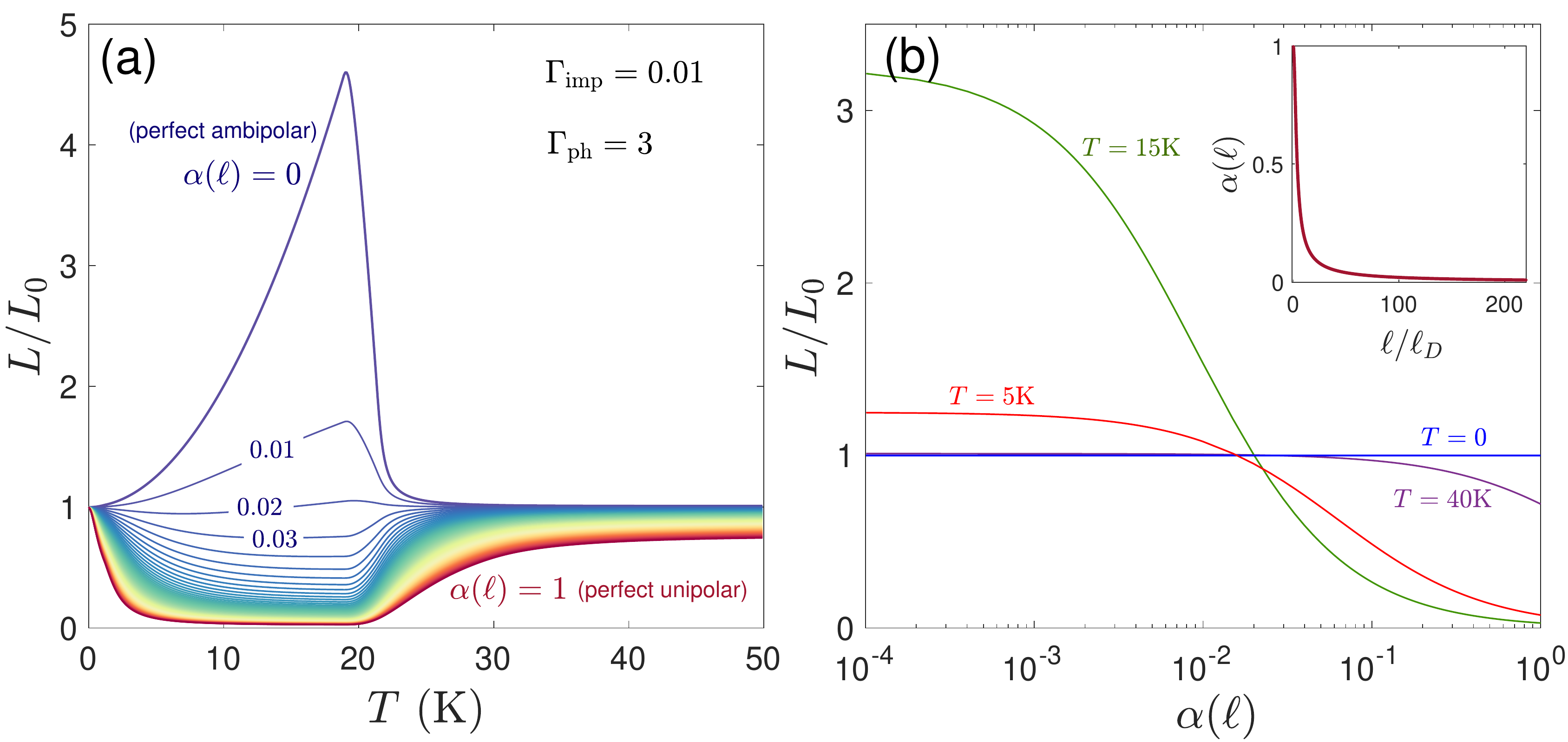}
\caption{
Lorenz ratio as a function of (a) temperature for different values of $\alpha(\ell)$  from 0 (perfect ambipolar regime) to 1 (perfect unipolar regime) with a step of  $0.01$ and (b) as a function of $\alpha(\ell)$ (in logarithmic scale) for different temperatures as labeled. The strength of the charge impurity- and phonon-limited resistivities were defined through the hydrodynamicity parameters $\Gamma_{\rm imp}^2=\rho_{\rm el,imp}/\rho_{\rm el}(T\gg T_F)$ and 
$\Gamma_{\rm ph}^2=\rho_{\rm el,ph}/\rho_{\rm el}(T\gg T_F)$,  respectively.  The inset in (b) shows $\alpha(\ell)$, Eq. (\ref{WeightFunction}), as a function of $\ell/\ell_D$.  The onset temperature for phonons is taken as $T_{\rm ph}=25$K. 
In the hydrodynamic regime, i.e. $ 0\lesssim T \lesssim 25$K in this figure, the Lorenz ratio deviates from its standard (impurity/phonon)-limited value $L_0$. Due to  the ambipolar constraint $j_e+j_h=0$, the violation of the Lorenz ratio is a large enhancement while with the unipolar situation ($j_e=j_h=0$) the Lorenz number is drastically reduced below $L_0$.} 
\label{fig:LL0}
\end{SCfigure*}

\section{Numerical Results and Discussion}
In this section we present numerical results for Eqs. (\ref{eq:theory_rho_t_result}), (\ref{eq:theory_Q_result}), and (\ref{eq:theory_WF_result}) for the thermal resistivity, the Seebeck coefficient, and  the Lorenz ratio, respectively. We assume the semimetal to be perfectly compensated, {\it i.e.} the number of electrons equals the number of holes. For the sake of clarity, we start by discussing the electrical and thermal resistivities in the ideal intrinsic limit, in which momentum-non-conserving interactions are completely neglected. These will be re-introduced later, in order to regularize the results pertaining to the Lorenz ratio. 

Our results are summarized in Fig.~\ref{fig:RthRel}. We plot the dimensionless electrical and thermal resistivities, $\rho_{\rm el}$ and $\rho_{\rm th}$, as  functions of temperature.  The thermal resistivities $\rho_{\rm th,uni}$ and $\rho_{\rm th, ambi}$, subject to  homogeneous unipolar and ambipolar boundary conditions,  are plotted as solid and dotted lines respectively.  As discussed above, the unipolar boundary condition sets electron and hole currents,  $j_{e}$ and $j_{h}$, individually  to zero.  This is equivalent to requiring that both the electric and the imbalance currents vanish.   On the other hand, the ambipolar boundary condition sets only the total current $j_e+j_h$ to zero.  Note that, as explained in the previous section, the distinction between unipolar and ambipolar boundary conditions does not apply to the electric resistivity. In Fig.~\ref{fig:RthRel} we see that, while the ambipolar thermal resistivity vanishes,  the unipolar thermal resistivity remains finite.   In the latter case the system is essentially equivalent to two independent electron and hole fluids with no electron-hole interactions. 

Since our model enjoys particle-hole symmetry, the thermal conductivity of electrons and holes are equal. Thus, $\rho_{\rm{th,uni}}=1/(2\kappa_{\rm e})$. We use the well-established kinetic-equation methods of Refs.~\cite{zarenia} and \cite{zareniaBLG} to calculate the intrinsic thermal conductivity $\kappa_{\rm e}$ of a single parabolic band at a fixed carrier density, defined by  the Fermi energy $\varepsilon_F$. These techniques can be viewed as a simplified version of the calculations performed in Ref.~\cite{Maslov2018}. We rely on a simpler Ansatz for the non-equilibrium distribution function~\cite{zarenia} and find
$\rho_{\rm{th,uni}}\simeq I_{\rm th}/(2D_{\rm th}^2)$, 
where $D_{\rm th}=\sum_{\boldsymbol{k}}(\partial f_{\boldsymbol{k}}/\partial \epsilon_{\boldsymbol{k}}) \boldsymbol{v}_{\boldsymbol{k}}\cdot \boldsymbol{v}_{\boldsymbol{k}} [(\epsilon_{\boldsymbol{k}}-\varepsilon_F)/k_BT)^2
\simeq 9\zeta(3)/(4\pi\hbar^2\beta)$ is the thermal  Drude weight, while $I_{\rm th}$ is the Coulomb collision integral projected onto the thermal channel.
%
In the degenerate Fermi liquid regime ($T\ll T_F$), $D_{\rm th}\sim T$ and $I_{\rm th}\sim T^4$ we find\footnote{For a system with a single parabolic band, the velocity $v_{\boldsymbol{k}}\sim \boldsymbol{k}$ coincides with the momentum zero mode, and therefore only the thermal moment of the collision integral, $I_{\rm th}$, associated with the relaxation of energy (thermal) currents, survives. This implies that, while the electric resistivity is exactly zero, the thermal resistivity remains finite. We make use of standard approximations for the Coulomb collision integral (screened interaction plus Fermi golden rule), previously used for graphene systems \cite{zarenia,zareniaBLG}, and find $I_{\rm th}$ to be
\be
I_{\rm th}=-\frac{1 }{4\pi(k_BT)^3}\sum_{\boldsymbol{q}}\int_{-\infty}^{\infty}d \omega\frac{|V(\boldsymbol{q},\omega)|^2}{\sinh^2(\hbar\omega/2k_BT)}[(\Im\Pi_1)^2-\Im\Pi_0\Im\Pi_2],
\ee
where $V(\qv,\omega)=v_q/|1-v_q\Pi_0(\qv,\omega)|$ is the screened electron-electron Coulomb interaction and $v_q=2\pi e^2/(\epsilon q)$. Here, $\epsilon$ is the dielectric constant that accounts for the surrounding medium as well as screening from remote bands. We set $\epsilon=1$ in our calculation. The response functions $\Pi_{n}(\qv,\omega)$ are defined as
\be
\Pi_n(\qv,\omega)=2\sum_{\boldsymbol{k}} \frac{(\tilde{\epsilon}_{\boldsymbol{k}} v_{\boldsymbol{k}} -\tilde{\epsilon}_{\boldsymbol{k}+\boldsymbol{q}}v_{\boldsymbol{k}+\boldsymbol{q}})^n(f^0_{\boldsymbol{k}} -f^0_{\boldsymbol{k}+\boldsymbol{q}})}{\epsilon_{\boldsymbol{k}} -\epsilon_{\boldsymbol{k}+\boldsymbol{q}}+\hbar\omega+i0^+} ,
\ee
where $\tilde\epsilon_{\boldsymbol{k}}=\epsilon_{\boldsymbol{k}}-\varepsilon_F$ is the band energy measured from the Fermi energy. At $T\ll T_F$ (degenerate regime), $\Pi_0$ is the well-known zero-temperature 2D Lindhard function \cite{vignalebook} and we find that for $\hbar\omega\ll \varepsilon_F$,  $[\Pi_1^2-\Pi_0\Pi_2]\sim \omega^6 $. This results in  $I_{\rm th}\sim T^4$.}
\be
\rho_{\rm{th,uni}}(T\ll T_F)\sim T^2\ln(T).
\ee
This behavior of the thermal resistivity is consistent with previous results obtained for degenerate Fermi liquid graphene~\cite{vignale}.
We find that $\rho_{\rm{th,uni}}$ peaks around $T\simeq T_F$ and decreases as $\sim \ln(T)/T^2$ for $T\gg T_F$ ({\it i.e.} in the non-degenerate regime), see blue solid curve in Fig. \ref{fig:RthRel}. 

Next, we study the the temperature dependence of the intrinsic electric resistivity $\rho_{\rm el}$ of the compensated semimetal. 
We distinguish the {\it Planckian} regime, in which there is only one energy scale
$k_BT$ ($T\gg T_F$), from the Fermi-liquid one, in which there are two energy scales, $k_BT$
and $\varepsilon_F$. The temperature dependence of $\rho_{\rm el}$ (in reduced units) is given by, 
\be
\rho_{{\rm el}} \sim \frac{1}{D_{\rm el}(T)\tau_{\rm eh}(T)},
\ee
where $\tau_{\rm eh}(T)$ is the electron-hole scattering rate and $D_{\rm el}(T)$ the Drude weight in the electric channel, which is defined as $D_{\rm el}=\sum_{\boldsymbol{k}}  \boldsymbol{v}_{\boldsymbol{k}}^2 (\partial f_{\boldsymbol{k}}/\partial \epsilon_{\boldsymbol{k}})$.
In the Planckian regime the maximum scattering rate allowed by the energy-time uncertainty principle \cite{Hartnoll2015} is $1/\tau_{\rm eh}(T)\sim k_BT/\hbar$ while in the Fermi liquid regime  $1/\tau_{\rm eh}(T)\sim (k_BT)^2/\hbar\varepsilon_F$. 
Similarly, in the Planckian regime the Drude weight $D_{\rm el}\propto k_BT$, whereas in the Fermi liquid regime it is independent of temperature. Hence, the intrinsic resistivity $\rho_{\rm el}$ is independent of temperature in the
Planckian regime, while it scales as $\sim T^2$ in the Fermi liquid regime. The red dashed curve in Fig. \ref{fig:RthRel} shows precisely these limiting behaviors. 

We now introduce momentum-non-conserving interactions and discuss the calculation of the Lorenz ratio of Eq.~(\ref{eq:theory_WF_result}). Momentum-non-conserving scattering is necessary to regularize results in the ambipolar limit: as it is clear from Fig.~\ref{fig:RthRel} and the definition~\ref{eq:theory_WF_result}, the ratio between the electrical and thermal resistivity ($\rho_{\rm th, ambi}$) would diverge if the contribution of disorder were neglected. To include disorder we consider the following simple but realistic model. At low temperatures impurities are the dominant disorder mechanism while, as temperature increases, electron-phonon scatterings become more important. 

The Drude resistivity due to scattering against impurities is here defined as $\rho_{\rm el,imp}=m^\ast/(n \tau_{\rm imp})$, where $m^\ast$ and $n$ are the electron effective mass and density, respectively. The impurity scattering rate $1/\tau_{\rm imp}$ is assumed to be independent of temperature for both short- and long-range impurities. 
Since the particle density $n$ and the effective masses are fixed in compensated semimetals,  once the electron and hole Fermi energies are set, the electric resistivity due to scattering against impurities, $\rho_{\rm el,imp}$, is independent of $T$.  

As the temperature increases, the resistivity due to collisions with phonons $\rho_{\rm el, ph}$ becomes the dominant contribution to the total electric resistivity. Above the Bloch-Gr\"uneisen temperature, $\rho_{\rm ph}$ increases linearly with $T$~\cite{Kasap2017}. We therefore posit the following model for the momentum-non-conserving scattering: $\rho_{\rm el,dis} (T\lesssim T_{\rm ph})\simeq \rho_{\rm el,imp}\sim \rm{Const.}$ and $\rho_{\rm el,dis} (T\gtrsim T_{\rm ph})\simeq \rho_{\rm el,ph}\sim T$, where $T_{\rm ph}$ is defined as the onset temperature at which phonons start to become the dominant scattering mechanism. 
Assuming that the Wiedemann-Franz law is satisfied when only momentum-non-conserving (electron-impurity or electron-phonon) processes are taken into account, we obtain in particular that the thermal resistivity of impurities in reduced units is $\rho_{\rm th,imp}=(3/\pi^2)\rho_{\rm el,imp}$.

Figures \ref{fig:LL0}a and \ref{fig:LL0}b show the results for the Lorenz ratio of Eq. (\ref{eq:theory_WF_result}) as a function of (a) temperature for different values of $\alpha(\ell)$ (corresponding to different sample lengths) and (b) as a function of $\alpha(\ell)$ for different temperatures. 
The onset temperature for phonon-dominated scattering is taken to be $T_{\rm ph}\simeq 25$ K.  We determine  the strength the charge impurity as well as phonon resistivities, respectively through the hydrodynamicity parameters $\Gamma_{\rm imp}^2=\rho_{\rm imp}/\rho_{\rm el}(T\gg T_F)$, and $\Gamma_{\rm ph}^2=\rho_{\rm ph}/\rho_{\rm el}(T\gg T_F)$ (i.e. the ratio of the charge impurity/phonon resistivities to the intrinsic Coulomb resistivity in the non-degenerate regime, $T\gg T_F$). For the results in Figs. \ref{fig:LL0}a and \ref{fig:LL0}b we have taken  $\Gamma_{\rm imp}=0.01$ and $\Gamma_{\rm ph}=3$.

We observe that the Wiedemann-Franz law is violated in two radically different ways, depending on whether we are in the ambipolar ($\ell\gg \ell_D$) or unipolar ($\ell\ll \ell_D$) limit. While in the former we observe a large enhancement of the Lorenz ratio, in the latter we observe a moderate reduction. The ambipolar limit is the situation realized in graphene systems (see Fig. \ref{fig:fig1}b), while the unipolar one occurs in compensated semimetals as WP$_2$ (see Fig. \ref{fig:fig1}c). We stress that the large enhancement of the Lorenz ratio cannot be explained without taking into account  strong electron-hole scattering in the electric conduction channel. 

\begin{figure}[h]
\centering 
\includegraphics[width=8cm]{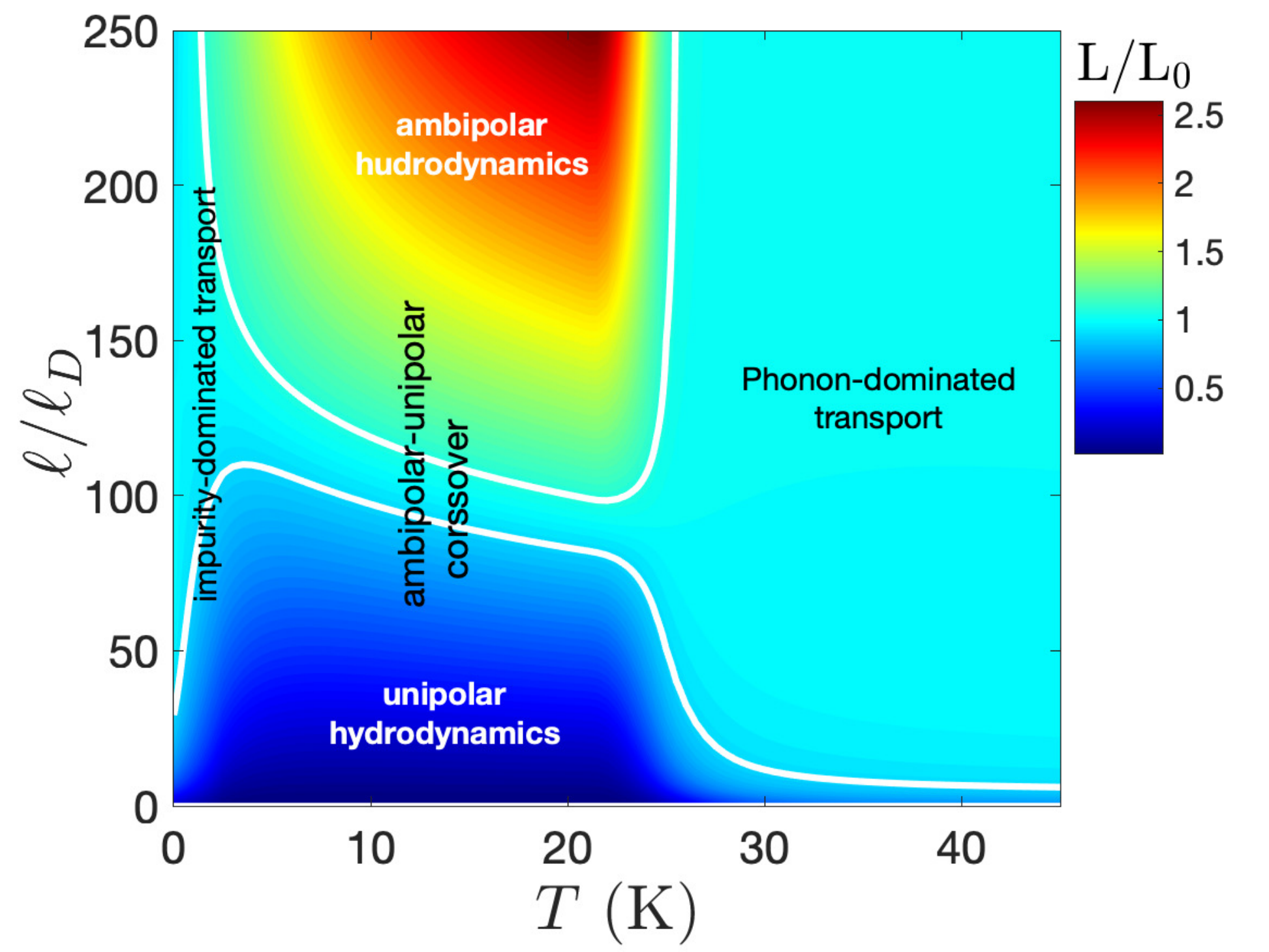}
\caption{2D plot of the Lorenz ratio as a function of $\ell/\ell_D$ (sample length scaled with the diffusion length) and $T$. Different transport regimes are indicated on the figure. The onset temperature of phonons $T_{\rm ph}=25$K and the hydrodynamicity parameters  $\Gamma_{\rm imp}=0.01$ and $\Gamma_{\rm ph}=3$ were chosen to be the same as in Fig. \ref{fig:LL0}. 
The enhancement of the Lorenz ratio (in the ambipolar hydrodynamic regime) is relevant for graphene systems and is a consequence of  electron-hole scattering, which selectively enhances the electric resistivity. The reduced Lorenz number (in the unipolar hydrodynamic regime) is relevant for compensated semimetals, where the electron and hole bands are well separated in momentum space.} 
\label{fig:phase}
\end{figure}

Fig. \ref{fig:phase} displays a 2D plot of $L/L_0$ as functions of temperature and scaled sample length ($\ell/\ell_D$). Based on the behavior of the Lorenz ratio we  identify a phase diagram of possible transport regimes in a charge-neutral system. When $T\to0$ as well as for temperatures $T\gtrsim T_{\rm ph}$, $L/L_0\to 1$ resulting from the disorder-limited transport in these regimes, i.e. impurity-dominated at $T\to0$ and phonon-dominated at $T\gtrsim T_{\rm ph}$). In the hydrodynamic regime ($0\lesssim T\lesssim T_{\rm ph}$), one can tune the WF ratio from an enhancement when $\ell\gg \ell_D$ (bipolar condition) to a reduction when $\ell\ll \ell_D$ (unipolar condition). 

Finally, we calculate the Seebeck coefficient according to Eq.~(\ref{eq:theory_Q_result}).
For a symmetric electron-hole system ($Q_e=-Q_h$), the Seebeck coefficient always vanishes at the charge neutrality point, both in the absence and in the presence of disorder. Note that in the intrinsic regime at perfect compensation, in which momentum-non-conserving processes are absent, $\rho_{\rm th, ambi} = 0$. This in turn implies that $\rho_{\rm th} = \alpha(\ell) \rho_{\rm th, uni}$ and $Q=Q_{\rm uni}$.  In this case, while the thermal resistivity depends on the system size, the Seebeck coefficient is independent of it.

\section{Outlook}

The breakdown of the Wiedemann-Franz law, which results in an anomalously large Lorenz ratio near the charge neutrality has been recognized to occur in clean graphene samples and has inspired a considerable amount of theoretical work. On the contrary, experiments in three-dimensional semimetals such as WP$_2$ and antimony show a radically different result. Although a phenomenology similar to that of graphene would naively be expected, a puzzling {\it reduced} Lorenz ratio is observed at low temperatures, when both electrons and holes form  degenerate Fermi liquids.

In this study, we have shown that this apparent contradiction is explained by the completely different transport situations realized in the two systems: truly ambipolar transport in graphene versus  two independent channels of unipolar transport in compensated semimetals.  In contrast to graphene and its bilayer, electron and hole pockets in compensated semimetals are well distanced in momentum space, resulting in a long recombination time. Since both electron and hole currents must separately vanish at the contacts, this results in a suppression of the bulk ambipolar conduction mechanism. Effectively, electrons and holes behave as two independent and decoupled Fermi liquid  throughout the channel  (a situation analogous to  unipolar transport). 

The violation of the Wiedemann-Franz law in both the ambipolar and unipolar transport regimes occurs in a temperature window in which the so-called {\it hydrodynamic} regime of transport is realized, {\it i.e.} when momentum-conserving collisions among particles constitute the dominant scattering mechanism (we note that in a Fermi liquid the WF law is satisfied when disorder scattering dominates). 
We have presented a simple theory for a general unipolar/ambipolar system and demonstrated a crossover  from ambipolar to unipolar conduction as a function of a weight function (related to the electron-hole recombination time) which naturally leads to a sample-size-dependent thermal conductivity as observed in experiments.  Although our theory has been presented for a two-dimensional system, the results would qualitatively remain valid for three-dimensional ones.

%
\section{Acknowledgments}
We thank Kamran Behnia and Joseph Heremans for useful discussions. M.Z. and G.V. were supported by the U.S. Department of Energy (Office of Science) under grant No. DE-FG02-05ER46203.  A.P. was supported by the European Commission under the EU Horizon 2020 MSCA-RISE-2019 programme (project 873028 HYDROTRONICS).

\appendix
\section*{Appendix}

\section{Explicitly accounting for momentum: the four-mode theory}
\label{app:Nm1_Nm1_matrix}
In this appendix we consider the general case in which the total momentum, the conserved mode of hydrodynamic conduction, does not coincide with either the electric, thermal or imbalance currents. Given its importance in determining the transport properties of the system, it is necessary to include it explicitly. In the resulting four-mode description, the total momentum is added to the list of currents flowing in the system. To stress the fact that it is a zero mode of the resulting $4\times 4$ resistivity matrix, {\it i.e.} a nontrivial vector with eigenvalue zero, we will call the total momentum $j_0$. A force $F_0$ that couples explicitly to it will also be included. 
The goal of this appendix is therefore to show how the $4\times 4$ resistivity matrix that connects the four currents to the four fields can be down-folded to obtain the $3\times 3$ matrix of Eq.~(\ref{eq:rho_split_def}). We will guide the reader through this process and show that the determinant of the resulting resistivity matrix vanishes, as stated in Sect.~\ref{sect:ambipolar_quantitative_theory}. This in turn implies that the zero mode, although not explicit, is still included in the three-mode theory. Therefore, no information about the physical implications of the conservation of momentum is lost in the down-folding process.

The derivation here parallels that given in Sect.~\ref{sect:ambipolar_quantitative_theory}. Fields and currents are now related by the $4\times 4$ resistivity matrix ${\hat \rho}$. Explicitly,
\be \label{eq:app_rho_split_def}
\left(
\begin{array}{c}
F_0
\\
\hline
\vect{F_{ns\delta}}
\end{array}
\right)
=
\left(
\begin{array}{c|c}
\rho_{00} & \tvect{\rho_0}
\\
\hline
\vect{\rho_0} & {\hat \rho}_{ns\delta}
\end{array}
\right)
\left(
\begin{array}{c}
j_0
\\
\hline
\vect{j_{ns\delta}}
\end{array}
\right)
~.
\ee
Here $\vect{j_{ns\delta}} = {}^t(j_n, j_s,j_\delta)$ and $\vect{F_{ns\delta}} = {}^t(-e E, -k_{\rm B}\partial_x T, F_\delta)$, ${\hat \rho}_{ns\delta}$ is a $3\times 3$ block, and $\vect{\rho_0}$ is a three-component vector. The determinant of the $4\times 4$ matrix in Eq.~(\ref{eq:app_rho_split_def}) is 
\be
\det({\hat \rho}) = \det({\hat \rho}_{ns\delta})\big[ \rho_{00} - \tvect{\rho_0} {\hat \rho}_{ns\delta}^{-1} \vect{\rho_0} \big]
~,
\ee
and it is equal to zero since there is at least one nontrivial vector with eigenvalue zero ({\it i.e.} the total momentum). Indeed, if the system is prepared in a state such that only $j_0\neq0$, the vector on the left-hand side of Eq.~(\ref{eq:app_rho_split_def}) is a null vector. By assumption, the only zero mode is the momentum, so the determinant of the $3\times 3$ block ${\hat \rho}_{ns\delta}$ is finite and such matrix is therefore invertible. Thus, it must be that
\be \label{eq:app_rho00_relation}
\rho_{00} = \tvect{\rho_0} {\hat \rho}_{ns\delta}^{-1} \vect{\rho_0}
~.
\ee
The property~(\ref{eq:app_rho00_relation}) plays a pivotal role in the following proof.

To down-fold the four-mode theory of Eq.~(\ref{eq:app_rho_split_def}) into the three-mode one of Eq.~(\ref{eq:rho_split_def}) we need to apply the boundary conditions on the momentum. Since no external field that couples specifically to the momentum is applied, we will set $F_0 = 0$, while the momentum $j_0$ is allowed to assume an arbitrary value. The latter is determined by the values of the electric and thermal currents and fields, as well as by the boundary conditions imposed on the imbalance current. The first line of Eq.~(\ref{eq:app_rho_split_def}) implies that
\be
j_0 = -\frac{\sprod{\rho_0}{j_{ns\delta}}}{\rho_{00}}
~.
\ee
When this relation is substituted into the last line of Eq.~(\ref{eq:app_rho_split_def}) we get
\be \label{eq:app_final_threemode_rho}
\vect{F_{ns\delta}} = \left[{\hat \rho}_{ns\delta} - \frac{\vect{\rho_0}\tvect{\rho_0}}{\rho_{00}} \right] \vect{j_{ns\delta}}
~.
\ee
The matrix in square brackets on the right-hand side of Eq.~(\ref{eq:app_final_threemode_rho}) is the $3\times3$ resistivity matrix of Eq.~(\ref{eq:rho_split_def}). We now prove that its determinant is zero. To do so, it is sufficient to show that there exists a nontrivial vector $\vect{j_{ns\delta}}$ such that, when the matrix acts on it, the result is exactly zero. It is easy to see that such vector is ${\hat \rho}_{ns\delta}^{-1}\vect{\rho_0}$. Indeed, using the property~(\ref{eq:app_rho00_relation}), we have
\ber
\vect{F_{ns\delta}} &=& \left[{\hat \rho}_{ns\delta} - \frac{\vect{\rho_0}\tvect{\rho_0}}{\rho_{00}} \right] {\hat \rho}_{ns\delta}^{-1}\vect{\rho_0}
\nn\\
&=& 
\vect{\rho_0} - \frac{\vect{\rho_0}\tvect{\rho_0}{\hat \rho}_{ns\delta}^{-1}\vect{\rho_0}}{\tvect{\rho_0} {\hat \rho}_{ns\delta}^{-1} \vect{\rho_0}} = 0
~,
\eer
which proves the assertion. Hence, the $3\times3$ resistivity matrix of Eq.~(\ref{eq:rho_split_def}) can be assumed to have determinant equal to zero and to implicitly retain the information about the conservation of momentum by electron-electron interactions.


\section{References}
\begin{thebibliography}{32}
\providecommand{\natexlab}[1]{#1}
\providecommand{\url}[1]{\texttt{#1}}
\expandafter\ifx\csname urlstyle\endcsname\relax
  \providecommand{\doi}[1]{doi: #1}\else
  \providecommand{\doi}{doi: \begingroup \urlstyle{rm}\Url}\fi

\bibitem[Bandurin et~al.(2016)Bandurin, Torre, Kumar, Ben~Shalom, Tomadin,
  Principi, Auton, Khestanova, Novoselov, Grigorieva, Ponomarenko, Geim, and
  Polini]{bandurin}
D.~A. Bandurin, I.~Torre, R.~Krishna Kumar, M.~Ben~Shalom, A.~Tomadin,
  A.~Principi, G.~H. Auton, E.~Khestanova, K.~S. Novoselov, I.~V. Grigorieva,
  L.~A. Ponomarenko, A.~K. Geim, and M.~Polini.
\newblock Negative local resistance caused by viscous electron backflow in
  graphene.
\newblock \emph{Science}, 351\penalty0 (6277):\penalty0 1055--1058, 2016.
\newblock ISSN 0036-8075.
\newblock \doi{10.1126/science.aad0201}.

\bibitem[Ghahari et~al.(2016)Ghahari, Xie, Taniguchi, Watanabe, Foster, and
  Kim]{ghahari}
Fereshte Ghahari, Hong-Yi Xie, Takashi Taniguchi, Kenji Watanabe, Matthew~S.
  Foster, and Philip Kim.
\newblock Enhanced thermoelectric power in graphene: Violation of the mott
  relation by inelastic scattering.
\newblock \emph{Phys. Rev. Lett.}, 116:\penalty0 136802, Mar 2016.
\newblock \doi{10.1103/PhysRevLett.116.136802}.
\newblock URL \url{https://link.aps.org/doi/10.1103/PhysRevLett.116.136802}.

\bibitem[Crossno et~al.(2016)Crossno, Shi, Wang, Liu, Harzheim, Lucas, Sachdev,
  Kim, Taniguchi, Watanabe, Ohki, and Fong]{crossno}
Jesse Crossno, Jing~K. Shi, Ke~Wang, Xiaomeng Liu, Achim Harzheim, Andrew
  Lucas, Subir Sachdev, Philip Kim, Takashi Taniguchi, Kenji Watanabe,
  Thomas~A. Ohki, and Kin~Chung Fong.
\newblock Observation of the dirac fluid and the breakdown of the
  wiedemann-franz law in graphene.
\newblock \emph{Science}, 2016.
\newblock ISSN 0036-8075.
\newblock \doi{10.1126/science.aad0343}.

\bibitem[Principi et~al.(2016)Principi, Vignale, Carrega, and
  Polini]{Principi_prb_2016}
Alessandro Principi, Giovanni Vignale, Matteo Carrega, and Marco Polini.
\newblock Bulk and shear viscosities of the two-dimensional electron liquid in
  a doped graphene sheet.
\newblock \emph{Phys. Rev. B}, 93:\penalty0 125410, Mar 2016.
\newblock \doi{10.1103/PhysRevB.93.125410}.
\newblock URL \url{https://link.aps.org/doi/10.1103/PhysRevB.93.125410}.

\bibitem[Narozhny et~al.(2015)Narozhny, Gornyi, Titov, Sch\"utt, and
  Mirlin]{Narozhny_prb_2015}
B.~N. Narozhny, I.~V. Gornyi, M.~Titov, M.~Sch\"utt, and A.~D. Mirlin.
\newblock Hydrodynamics in graphene: Linear-response transport.
\newblock \emph{Phys. Rev. B}, 91:\penalty0 035414, Jan 2015.
\newblock \doi{10.1103/PhysRevB.91.035414}.
\newblock URL \url{https://link.aps.org/doi/10.1103/PhysRevB.91.035414}.

\bibitem[Briskot et~al.(2015)Briskot, Sch\"utt, Gornyi, Titov, Narozhny, and
  Mirlin]{Briskot_prb_2015}
U.~Briskot, M.~Sch\"utt, I.~V. Gornyi, M.~Titov, B.~N. Narozhny, and A.~D.
  Mirlin.
\newblock Collision-dominated nonlinear hydrodynamics in graphene.
\newblock \emph{Phys. Rev. B}, 92:\penalty0 115426, Sep 2015.
\newblock \doi{10.1103/PhysRevB.92.115426}.
\newblock URL \url{https://link.aps.org/doi/10.1103/PhysRevB.92.115426}.

\bibitem[Fritz et~al.(2008)Fritz, Schmalian, M\"uller, and Sachdev]{Fritz_2008}
Lars Fritz, J\"org Schmalian, Markus M\"uller, and Subir Sachdev.
\newblock Quantum critical transport in clean graphene.
\newblock \emph{Phys. Rev. B}, 78:\penalty0 085416, Aug 2008.
\newblock \doi{10.1103/PhysRevB.78.085416}.
\newblock URL \url{https://link.aps.org/doi/10.1103/PhysRevB.78.085416}.

\bibitem[M\"uller et~al.(2008)M\"uller, Fritz, and Sachdev]{Muller_2008}
Markus M\"uller, Lars Fritz, and Subir Sachdev.
\newblock Quantum-critical relativistic magnetotransport in graphene.
\newblock \emph{Phys. Rev. B}, 78:\penalty0 115406, Sep 2008.
\newblock \doi{10.1103/PhysRevB.78.115406}.
\newblock URL \url{https://link.aps.org/doi/10.1103/PhysRevB.78.115406}.

\bibitem[Xie and Foster(2016)]{foster}
Hong-Yi Xie and Matthew~S. Foster.
\newblock Transport coefficients of graphene: Interplay of impurity scattering,
  coulomb interaction, and optical phonons.
\newblock \emph{Phys. Rev. B}, 93:\penalty0 195103, May 2016.
\newblock \doi{10.1103/PhysRevB.93.195103}.
\newblock URL \url{https://link.aps.org/doi/10.1103/PhysRevB.93.195103}.

\bibitem[Svintsov et~al.(2013)Svintsov, Vyurkov, Ryzhii, and
  Otsuji]{svintsov2013}
D.~Svintsov, V.~Vyurkov, V.~Ryzhii, and T.~Otsuji.
\newblock Hydrodynamic electron transport and nonlinear waves in graphene.
\newblock \emph{Phys. Rev. B}, 88:\penalty0 245444, Dec 2013.
\newblock \doi{10.1103/PhysRevB.88.245444}.
\newblock URL \url{https://link.aps.org/doi/10.1103/PhysRevB.88.245444}.

\bibitem[Svintsov(2018)]{svintsov2018}
D.~Svintsov.
\newblock Hydrodynamic-to-ballistic crossover in dirac materials.
\newblock \emph{Phys. Rev. B}, 97:\penalty0 121405, Mar 2018.
\newblock \doi{10.1103/PhysRevB.97.121405}.
\newblock URL \url{https://link.aps.org/doi/10.1103/PhysRevB.97.121405}.

\bibitem[Ho et~al.(2018)Ho, Yudhistira, Chakraborty, and Adam]{shaffique}
Derek Y.~H. Ho, Indra Yudhistira, Nilotpal Chakraborty, and Shaffique Adam.
\newblock Theoretical determination of hydrodynamic window in monolayer and
  bilayer graphene from scattering rates.
\newblock \emph{Phys. Rev. B}, 97:\penalty0 121404, Mar 2018.
\newblock \doi{10.1103/PhysRevB.97.121404}.
\newblock URL \url{https://link.aps.org/doi/10.1103/PhysRevB.97.121404}.

\bibitem[Principi and Vignale(2015{\natexlab{a}})]{Principi_conductivity}
Alessandro Principi and Giovanni Vignale.
\newblock Intrinsic charge and spin conductivities of doped graphene in the
  fermi-liquid regime.
\newblock \emph{Phys. Rev. B}, 91:\penalty0 205423, May 2015{\natexlab{a}}.
\newblock \doi{10.1103/PhysRevB.91.205423}.
\newblock URL \url{https://link.aps.org/doi/10.1103/PhysRevB.91.205423}.

\bibitem[Lucas and Das~Sarma(2018)]{lucasWF}
Andrew Lucas and Sankar Das~Sarma.
\newblock Electronic hydrodynamics and the breakdown of the wiedemann-franz and
  mott laws in interacting metals.
\newblock \emph{Phys. Rev. B}, 97:\penalty0 245128, Jun 2018.
\newblock \doi{10.1103/PhysRevB.97.245128}.
\newblock URL \url{https://link.aps.org/doi/10.1103/PhysRevB.97.245128}.

\bibitem[Lucas and Hartnoll(2018)]{lucasKT}
Andrew Lucas and Sean~A. Hartnoll.
\newblock Kinetic theory of transport for inhomogeneous electron fluids.
\newblock \emph{Phys. Rev. B}, 97:\penalty0 045105, Jan 2018.
\newblock \doi{10.1103/PhysRevB.97.045105}.
\newblock URL \url{https://link.aps.org/doi/10.1103/PhysRevB.97.045105}.

\bibitem[Zarenia et~al.(2019{\natexlab{a}})Zarenia, Principi, and
  Vignale]{zarenia}
Mohammad Zarenia, Alessandro Principi, and Giovanni Vignale.
\newblock Disorder-enabled hydrodynamics of charge and heat transport in
  monolayer graphene.
\newblock \emph{2D Materials}, 6\penalty0 (3):\penalty0 035024, may
  2019{\natexlab{a}}.
\newblock \doi{10.1088/2053-1583/ab1ad9}.
\newblock URL \url{https://doi.org/10.1088%2F2053-1583%2Fab1ad9}.

\bibitem[Zarenia et~al.(2019{\natexlab{b}})Zarenia, Smith, Principi, and
  Vignale]{zareniaBLG}
Mohammad Zarenia, Thomas~Benjamin Smith, Alessandro Principi, and Giovanni
  Vignale.
\newblock Breakdown of the wiedemann-franz law in $ab$-stacked bilayer
  graphene.
\newblock \emph{Phys. Rev. B}, 99:\penalty0 161407, Apr 2019{\natexlab{b}}.
\newblock \doi{10.1103/PhysRevB.99.161407}.
\newblock URL \url{https://link.aps.org/doi/10.1103/PhysRevB.99.161407}.

\bibitem[Zarenia et~al.(2020{\natexlab{a}})Zarenia, Yudhistira, Adam, and
  Vignale]{zareniaTwisted}
Mohammad Zarenia, Indra Yudhistira, Shaffique Adam, and Giovanni Vignale.
\newblock Enhanced hydrodynamic transport in near magic angle twisted bilayer
  graphene.
\newblock \emph{Phys. Rev. B}, 101:\penalty0 045421, Jan 2020{\natexlab{a}}.
\newblock \doi{10.1103/PhysRevB.101.045421}.
\newblock URL \url{https://link.aps.org/doi/10.1103/PhysRevB.101.045421}.

\bibitem[Zarenia et~al.(2020{\natexlab{b}})Zarenia, Adam, and
  Vignale]{zareniaCollapse}
Mohammad Zarenia, Shaffique Adam, and Giovanni Vignale.
\newblock Temperature collapse of the electric conductivity in bilayer
  graphene.
\newblock \emph{Phys. Rev. Research}, 2:\penalty0 023391, Jun
  2020{\natexlab{b}}.
\newblock \doi{10.1103/PhysRevResearch.2.023391}.
\newblock URL \url{https://link.aps.org/doi/10.1103/PhysRevResearch.2.023391}.

\bibitem[Jaoui et~al.(2018)Jaoui, Fauqu\'e, Rischau, Subedi, Fu, Gooth, Kumar,
  S\"u$\beta$, Maslov, Felser, and Behnia]{Jaoui2018}
Alexandre Jaoui, B.~Fauqu\'e, Carl~Willem Rischau, Alaska Subedi, Chenguang Fu,
  Johannes Gooth, Nitesh Kumar, Vicky S\"u$\beta$, Dmitrii~L. Maslov, Claudia
  Felser, and Kamran Behnia.
\newblock Departure from the wiedemann-franz law in wp2 driven by mismatch in
  t-square resistivity prefactors.
\newblock \emph{npj Quantum Materials}, 3\penalty0 (1):\penalty0 64, Dec 2018.
\newblock ISSN 2397-4648.
\newblock \doi{10.1038/s41535-018-0136-x}.
\newblock URL \url{https://doi.org/10.1038/s41535-018-0136-x}.

\bibitem[Gooth et~al.(2018)Gooth, Menges, Kumar, S\"{u}$\beta$, Shekhar, Sun,
  Drechsler, Zierold, Felser, and Gotsmann]{Gooth2018}
J.~Gooth, F.~Menges, N.~Kumar, V.~S\"{u}$\beta$, C.~Shekhar, Y.~Sun,
  U.~Drechsler, R.~Zierold, C.~Felser, and B.~Gotsmann.
\newblock Thermal and electrical signatures of a hydrodynamic electron fluid in
  tungsten diphosphide.
\newblock \emph{Nature Communications}, 9\penalty0 (1):\penalty0 4093, Oct
  2018.
\newblock ISSN 2041-1723.
\newblock \doi{10.1038/s41467-018-06688-y}.
\newblock URL \url{https://doi.org/10.1038/s41467-018-06688-y}.

\bibitem[Principi and Vignale(2015{\natexlab{b}})]{vignale}
Alessandro Principi and Giovanni Vignale.
\newblock Violation of the wiedemann-franz law in hydrodynamic electron
  liquids.
\newblock \emph{Phys. Rev. Lett.}, 115:\penalty0 056603, Jul
  2015{\natexlab{b}}.
\newblock \doi{10.1103/PhysRevLett.115.056603}.
\newblock URL \url{https://link.aps.org/doi/10.1103/PhysRevLett.115.056603}.

\bibitem[Jaoui et~al.(2020)Jaoui, Fauqué, and Behnia]{behnia}
Alexandre Jaoui, Benoît Fauqué, and Kamran Behnia.
\newblock Thermal resistivity due to electron viscosity in bulk antimony, 2020.

\bibitem[Gallo et~al.(1963)Gallo, Chandrasekhar, and Sutter]{Gallo63}
C.~F. Gallo, B.~S. Chandrasekhar, and P.~H. Sutter.
\newblock Transport properties of bismuth single crystals.
\newblock \emph{Journal of Applied Physics}, 34\penalty0 (1):\penalty0
  144--152, 1963.
\newblock \doi{10.1063/1.1729056}.
\newblock URL \url{https://doi.org/10.1063/1.1729056}.

\bibitem[Uher and Goldsmid(1974)]{Uher74}
C~Uher and H.~J. Goldsmid.
\newblock Separation of the electronic and lattice thermal conductivities in
  bismuth crystals.
\newblock \emph{Phys. Stat. Sol.}, 65:\penalty0 765, 1974.

\bibitem[Li and Maslov(2018)]{Maslov2018}
Songci Li and Dmitrii~L. Maslov.
\newblock Lorentz ratio of a compensated metal.
\newblock \emph{Phys. Rev. B}, 98:\penalty0 245134, Dec 2018.
\newblock \doi{10.1103/PhysRevB.98.245134}.
\newblock URL \url{https://link.aps.org/doi/10.1103/PhysRevB.98.245134}.

\bibitem[Pines and Nozi{\`e}res(1966)]{pines}
D.~Pines and P.~Nozi{\`e}res.
\newblock \emph{The Theory of Quantum Liquids: Normal Fermi liquids}.
\newblock W.A. Benjamin, 1966.
\newblock URL \url{https://books.google.com.sg/books?id=GP1QAAAAMAAJ}.

\bibitem[Streetman(1990)]{Streetman}
Ben~G. Streetman.
\newblock \emph{Solid State Electronic Devices}.
\newblock Prentice-Hall, Inc., USA, 1990.
\newblock ISBN 0138229414.

\bibitem[Chen et~al.(2019)Chen, Li, Zhao, Zhou, and Zhu]{Chen2019}
Yuzhong Chen, Yujie Li, Yida Zhao, Hongzhi Zhou, and Haiming Zhu.
\newblock Highly efficient hot electron harvesting from graphene before
  electron-hole thermalization.
\newblock \emph{Science Advances}, 5\penalty0 (11), 2019.
\newblock \doi{10.1126/sciadv.aax9958}.
\newblock URL \url{https://advances.sciencemag.org/content/5/11/eaax9958}.

\bibitem[Giuliani and Vignale(2005)]{vignalebook}
Gabriele Giuliani and Giovanni Vignale.
\newblock \emph{Quantum Theory of the Electron Liquid}.
\newblock Cambridge University Press, 2005.
\newblock \doi{10.1017/CBO9780511619915}.

\bibitem[Hartnoll(2015)]{Hartnoll2015}
Sean~A. Hartnoll.
\newblock Theory of universal incoherent metallic transport.
\newblock \emph{Nature Physics}, 11\penalty0 (1):\penalty0 54--61, Jan 2015.
\newblock ISSN 1745-2481.
\newblock \doi{10.1038/nphys3174}.
\newblock URL \url{https://doi.org/10.1038/nphys3174}.

\bibitem[Kasap et~al.(2017)Kasap, Koughia, and Ruda]{Kasap2017}
Safa Kasap, Cyril Koughia, and Harry~E. Ruda.
\newblock \emph{Electrical Conduction in Metals and Semiconductors}, pages
  1--1.
\newblock Springer International Publishing, Cham, 2017.
\newblock ISBN 978-3-319-48933-9.
\newblock \doi{10.1007/978-3-319-48933-9_2}.
\newblock URL \url{https://doi.org/10.1007/978-3-319-48933-9_2}.

\end{thebibliography}
\end{document}